\documentclass{jpsj-suppl}
\usepackage{txfonts} 
\usepackage{caption}

\title{Measurement of Electron Neutrino Charged-Current Inclusive Cross Section in 1$-$3 GeV energy region
with the NO$\nu$A Near Detector}

\author{Xuebing Bu}

\inst{Fermilab}

\email{xbbu@fnal.gov}

\recdate{January 4, 2015}

\abst{
We present a measurement of the electron neutrino charged-current inclusive cross section
per nucleon with a data sample corresponding to $2.6 \times 10^{20}$ protons-on-target collected
by NO$\nu$A near detector at Fermilab.
We measured the cross section in four energy bins from 1 GeV to 3 GeV with half GeV per bin.
}

\kword{nova, nue, cross section}

\begin{document}
\maketitle

\section{Introduction}
As summarized in Ref. \cite{Xnu}, there are few $\nu_{e}$ cross section
measurements.
LSND provided a measurement in the low energy region up to 55 MeV \cite{lsnd}.
The first measurement at GeV scale was from the Gargamelle experiment~\cite{gargamelle}
based on 200 selected data events.
The T2K experiment measured the cross section at neutrino energy
$E_{\nu_{e}} \approx 1$ GeV~\cite{t2k} using 315 selected data events.
Most recently, MINERvA has published a result using approximately 2100 candidate
events in the range (0.5-10 GeV) with average purity of 50\%~\cite{minerva-nue}.

NO$\nu$A~\cite{nova} is a long baseline neutrino experiment with 
two functionally equivalent detectors,
to study neutrino oscillations, neutrino interaction cross sections,
and new phenomena searches.
NO$\nu$A detectors are low-Z and highly active tracking calorimeters, designed to 
efficiently identify $\nu_{e}$ interactions.
The NO$\nu$A near detector is located approximately 14.6 mrad off the axis
of the NuMI neutrino beam~\cite{beam}, with a $\nu_{e}$ flux covering a broad range of energies.
In this proceeding, we present the measurement of electron neutrino charged-current inclusive 
cross section per nucleon in bins of neutrino energy from 1~GeV to 3~GeV,
using data corresponding to $2.6 \times 10^{20}$ protons-on-target
collected by NO$\nu$A near detector between November 2014 and June 2015.

\section{The NO$\nu$A near detector}
NO$\nu$A near detector (ND) is located about 100~m underground at Fermilab,
and 1~km from the source of NuMI neutrino beam.
That beam is produced from the decay of charged pions
and kaons generated by 120~GeV proton collisions on a graphite target.
The ND is positioned at an angle of 14.6 mrad relative to the beam axis
direction and composed of 214 layers of cells of extruded PVC~\cite{nova-pvc},
with 3.9~cm~$\times$~6.6~cm in area and 4~m in length.
There is a muon catcher, located at the end of the fully active detector,
composed by 22 layers of PVC cells and 10 layers steel plates with 10~cm thickness,
where each steel plate is put in between two layers of PVC cells. 
The muon catcher is $\frac{2}{3}$ the height of the fully active detector.
Each PVC cell contains single looped wavelength-shifting fiber and 
is filled with liquid scintillator pseudocumene~\cite{nova-scintillator},
where the scintillation light is transported
to one pixel of a 32-pixel Hamamatsu avalanche photodiode.
These planes of PVC cells are layered in orthogonal views
to allow for 3-dimensional reconstruction of neutrino interactions.
Overall, NO$\nu$A ND consists of 20k readout channels,
and weights 300 ton.

In this analysis, only fully active detector is used,
which has a total mass of 190 ton.
In the fiducial region of the detector, the liquid scintillator
comprises 63\% of the detector mass.
The ND is dominated by Carbon, Chlorine, and Hydrogen (see Table~\ref{detectormass}).
The total number of nucleons in the fiducial region is $2.7 \times 10^{31}$.

\begin{table}[tbh]
\centering
\caption{Mass weight for NO$\nu$A near detector components.}
\captionsetup{justification=centering}
\label{detectormass}
\begin{tabular}{cccccc}
\hline
C12 & Cl35 & H1 & Ti48 & O16 & others \\
66.8\% & 16.4\% & 10.5\% & 3.3\% & 2.6\% & 0.4\% \\
\hline
\end{tabular}
\end{table}

\section{Simulation and Calibration}
\begin{figure}[tbh]
\centering
\includegraphics[keepaspectratio,width=.4\textwidth]{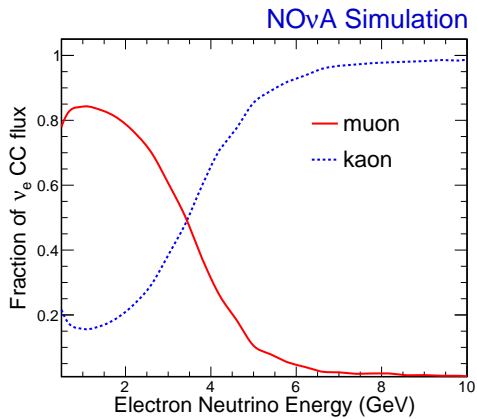}
\caption{Fraction of $\nu_{e}$ flux as a function of $\nu_{e}$ energy 
from muons and kaons.}
\label{nue-flux}
\end{figure}

The neutrino beam flux is simulated using the 
{\sc fluka}~\cite{fluka}, interfaced with a {\sc geant}4~\cite{geant} 
geometry with {\sc flugg}~\cite{flugg}
to model the interactions of protons with the target.
The neutrino beam is dominated by $\nu_{\mu}/\bar{\nu}_{\mu}$
with a $\approx$2\% intrinsic beam $\nu_{e}+\bar{\nu}_{e}$ component. 
These intrinsic beam $\nu_{e}$ dominantly come from the decay
of $\mu^{+}$ produced by pions' decay and from charged and neutral kaons.
The $\nu_{e}$ from muon decays contribute most 
in low energy region ($<3$ GeV), while above that energy the
$\nu_{e}$ comes primarily from kaon decays.
The fraction of contribution to $\nu_{e}$ flux as a function
of $\nu_{e}$ energy is shown in Fig.~\ref{nue-flux}.

The neutrino interactions inside and outside the detector
are simulated using {\sc genie}~\cite{genie},
and {\sc geant} is used to track the resulting particles and 
corresponding energy depositions.
Experiment-specific simulations~\cite{nova-sim} are used to model the capture
of the scintillation photons in the wavelength-shifting fibers
and the response of the readout electronics.

Cosmic ray muons are used as standard candles to calibrate
the detector response.
In addition, the energy spectrum of Michel electrons
and the $\pi^{0}$ mass peak, as well as the 
muons from neutrino interactions in near detector are used
as the cross-checks.
The detector response is simulated well;
all results agree within 5\%.

\section{Reconstruction and event selection}
The calibrated cell hits are grouped together to form
the basic reconstructed clusters based on
the time and space to construct the neutrino candidates~\cite{slice1,slice2}.
Each basic cluster corresponds to one neutrino interaction.
Within these basic clusters, the particle paths are found,
and the corresponding interactions of these paths are used as seeds
to find the neutrino vertex.
Furthermore, this event vertex is used as seed to form electromagnetic showers
using the Fuzzy-clustering algorithm~\cite{prong1,prong2}.

The $\nu_{e}$ CC candidates are selected by requiring the
most energetic EM shower (primary shower) to be contained, with energy less than 3.5~GeV.
In addition, the shower length is required to be greater than 150~cm
but less than 500~cm, and the fraction of MIP hits to be less than 35\%.
A likelihood-based selector is required to be greater than 0.2,
which was developed using an artificial neural network by combining the likelihood
differences of longitudinal and transverse shower energy deposition
among different particle hypotheses, and other topological
variables.

\begin{figure}[tbh]
\centering
\includegraphics[keepaspectratio,width=.4\textwidth]{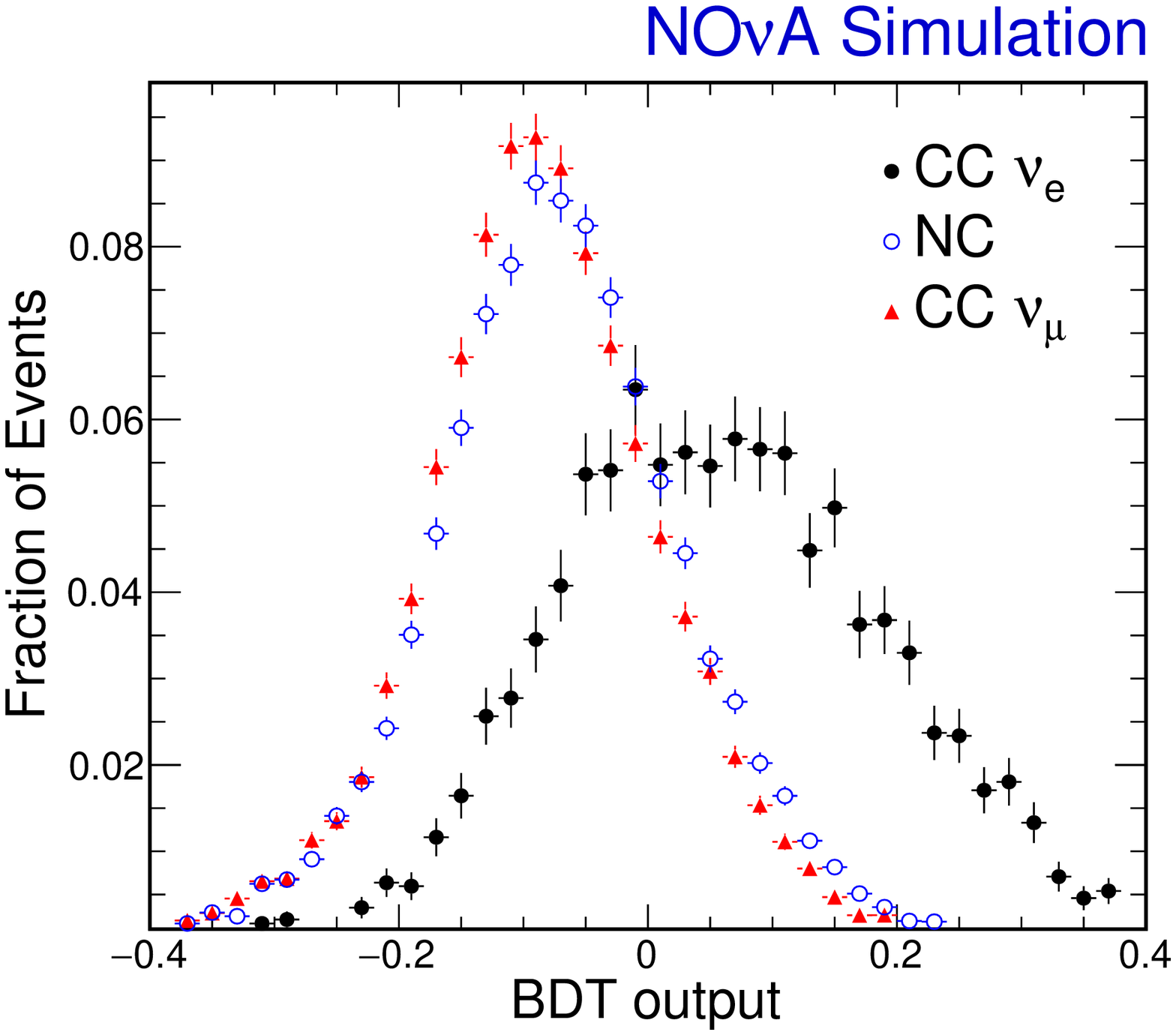}
\caption{Normalized BDT output spectrum from $\nu_{e}$ CC signal, and $\nu_{\mu}$ and NC backgrounds.}
\label{bdtoutput-mc}
\end{figure}

\begin{figure}[tbh]
\centering
\includegraphics[keepaspectratio,width=.6\textwidth]{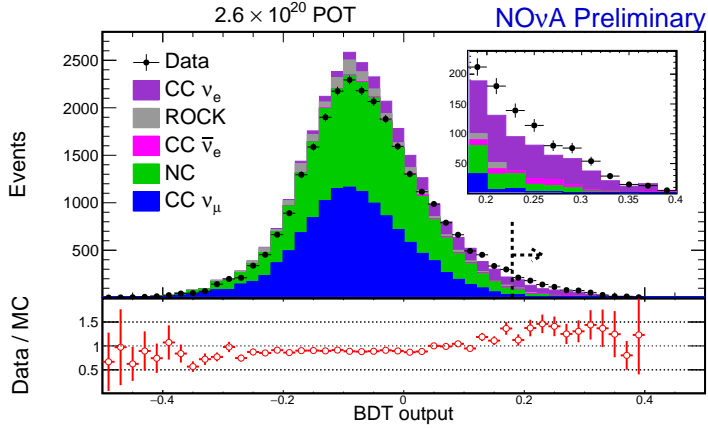}
\caption{BDT output spectrum from data, $\nu_{e}$ CC signal, and backgrounds.}
\label{bdtoutput-data}
\end{figure}

To reduce background further, a Boosted Decision Trees (BDT)~\cite{bdt-ref} discriminant is defined,
combining the fraction of MIP hits in the second most energetic shower,
the fraction of energy in transverse road around the primary shower core by $\pm$4~cm,
the maximal fraction of energy in 6-continuous planes along beam direction,
the fraction of energy in first 10 planes, and the fraction of energy
in secondary, third, and fourth plane of primary shower.
The shapes of BDT output, normalized to unity area and obtained after all 
event selection criteria, are shown in Fig.~\ref{bdtoutput-mc},
exhibiting a significant discrimination between $\nu_{e}$ CC signal, and $\nu_{\mu}$ and NC backgrounds.
The BDT output distributions for the data, signal, and backgrounds~(see Section~\ref{sect:bkg}) 
are shown
in Fig.~\ref{bdtoutput-data}, the BDT output $>$ 0.18 is used in this analysis to extract
the cross section.
After imposing all selection requirements,
917 $\nu_{e}$ CC candidates are selected in data, with an expected purity of 65\%.

\begin{figure}[tbh]
\centering
\includegraphics[keepaspectratio,width=.4\textwidth]{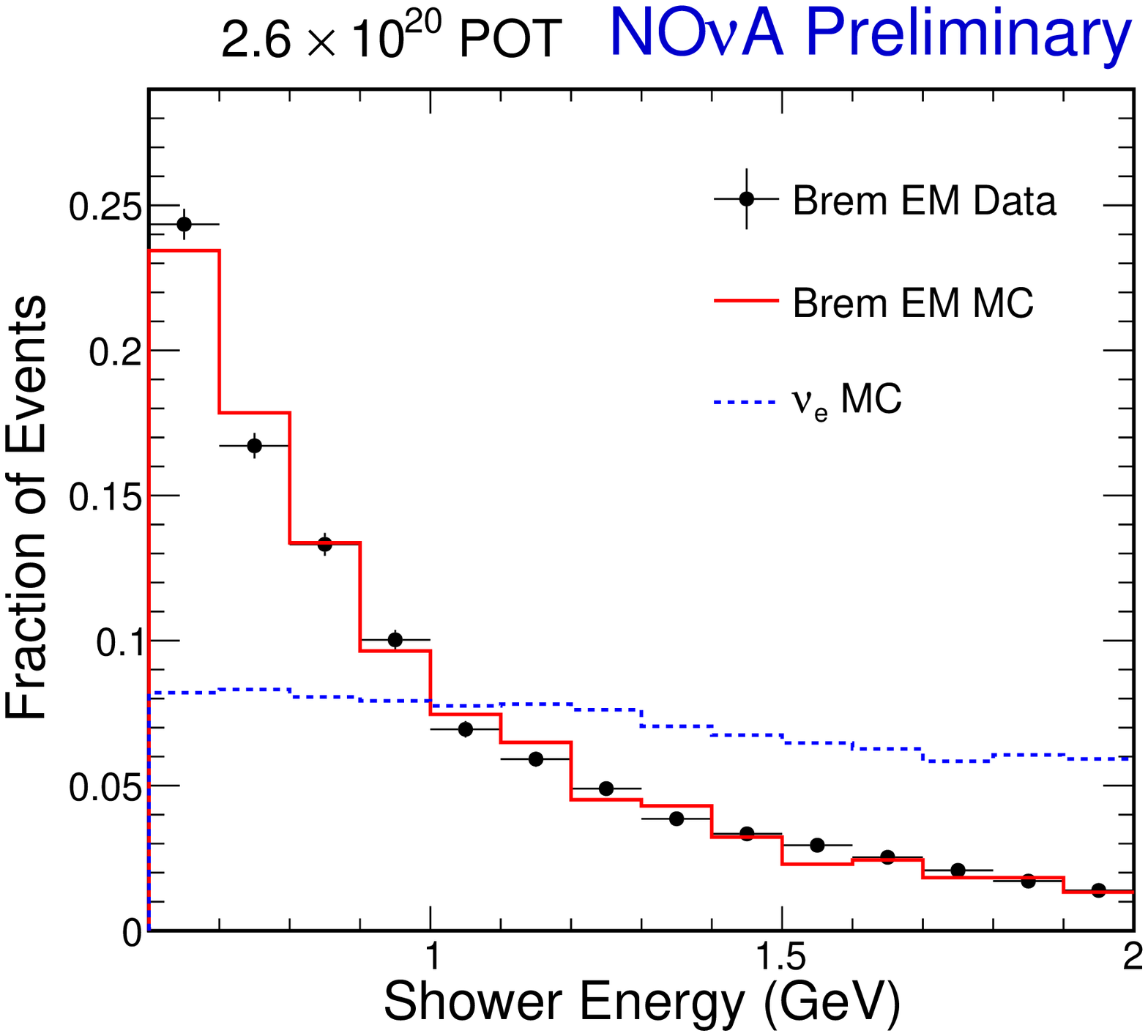}
\includegraphics[keepaspectratio,width=.4\textwidth]{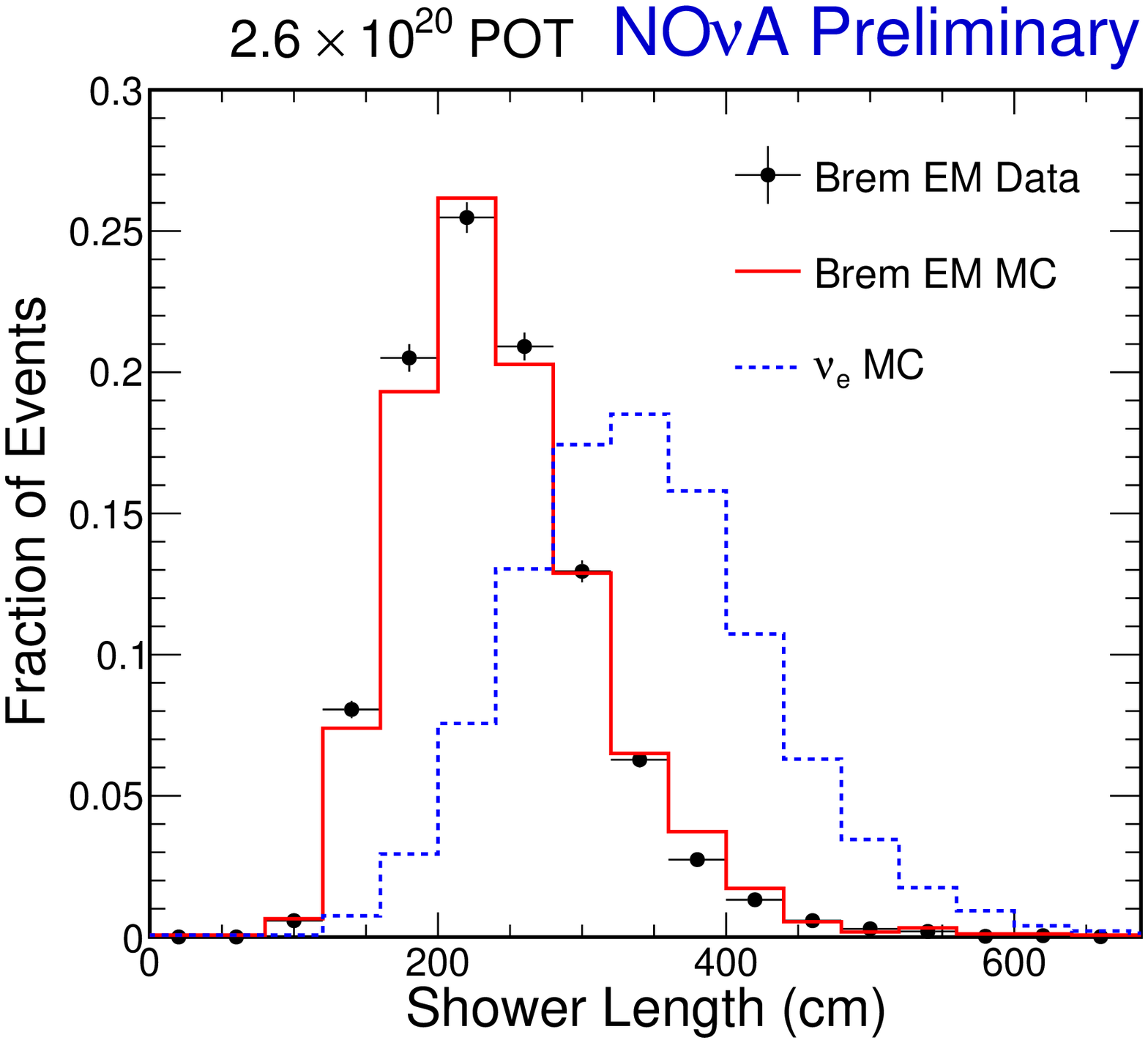}
\caption{Normalized shower energy (left plot) and length (right plot) distributions
from electromagnetic showers of muon bremsstrahlung in data and MC, and $\nu_{e}$ CC signal.}
\label{shower-shape}
\end{figure}

\begin{figure}[tbh]
\centering
\includegraphics[keepaspectratio,width=.4\textwidth]{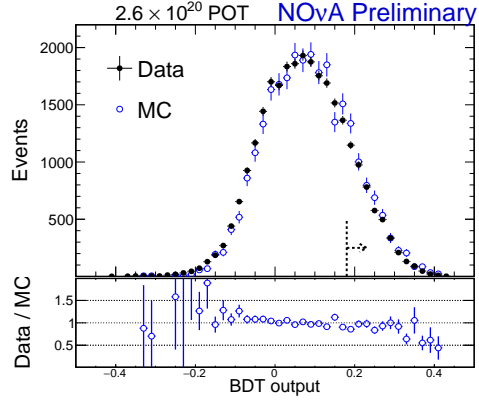}
\caption{BDT output spectrum from electromagnetic showers of muon bremsstrahlung in data and MC.}
\label{bdtoutput-bremEM}
\end{figure}

To validate the performance of EM shower reconstruction and the BDT discriminant, we select
the electromagnetic showers from muon bremsstrahlung in data.
While most muons produced in neutrino interactions and traversing the detector
usually leave a long, isolated track, some produce bremsstrahlung electromagnetic showers.
We select those events, remove the muon activity based on the energy deposition of MIP hits,
and reconstruct the electromagnetic showers from the remaining activity.
The shapes of the shower energy and length, normalized to unity area,
for electromagnetic showers from muon bremsstrahlung in data and monte carlo simulation (MC), 
and $\nu_{e}$ CC signal are shown in Fig.~\ref{shower-shape}.
There is good agreement between data and MC, and
these bremsstrahlung EM showers cover similar kinematic regions as 
the EM showers from the $\nu_{e}$ CC signal.
The BDT discriminant distributions for those bremsstrahlung electromagnetic showers
are shown in Fig.~\ref{bdtoutput-bremEM}, showing good agreement between data and MC.
 
\section{Background subtraction}
\label{sect:bkg}
\begin{figure}[tbh]
\centering
\includegraphics[keepaspectratio,width=.4\textwidth]{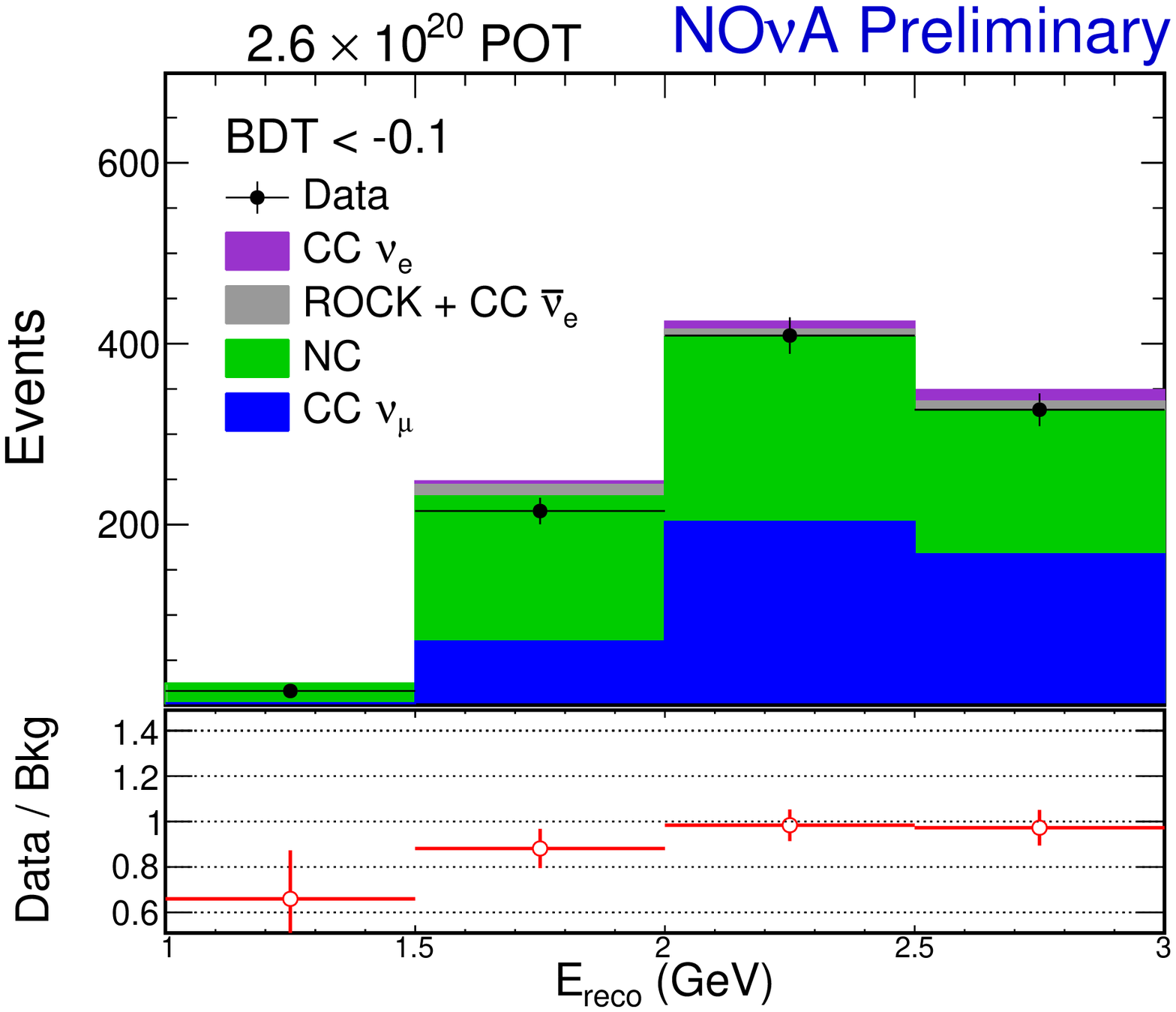}
\includegraphics[keepaspectratio,width=.4\textwidth]{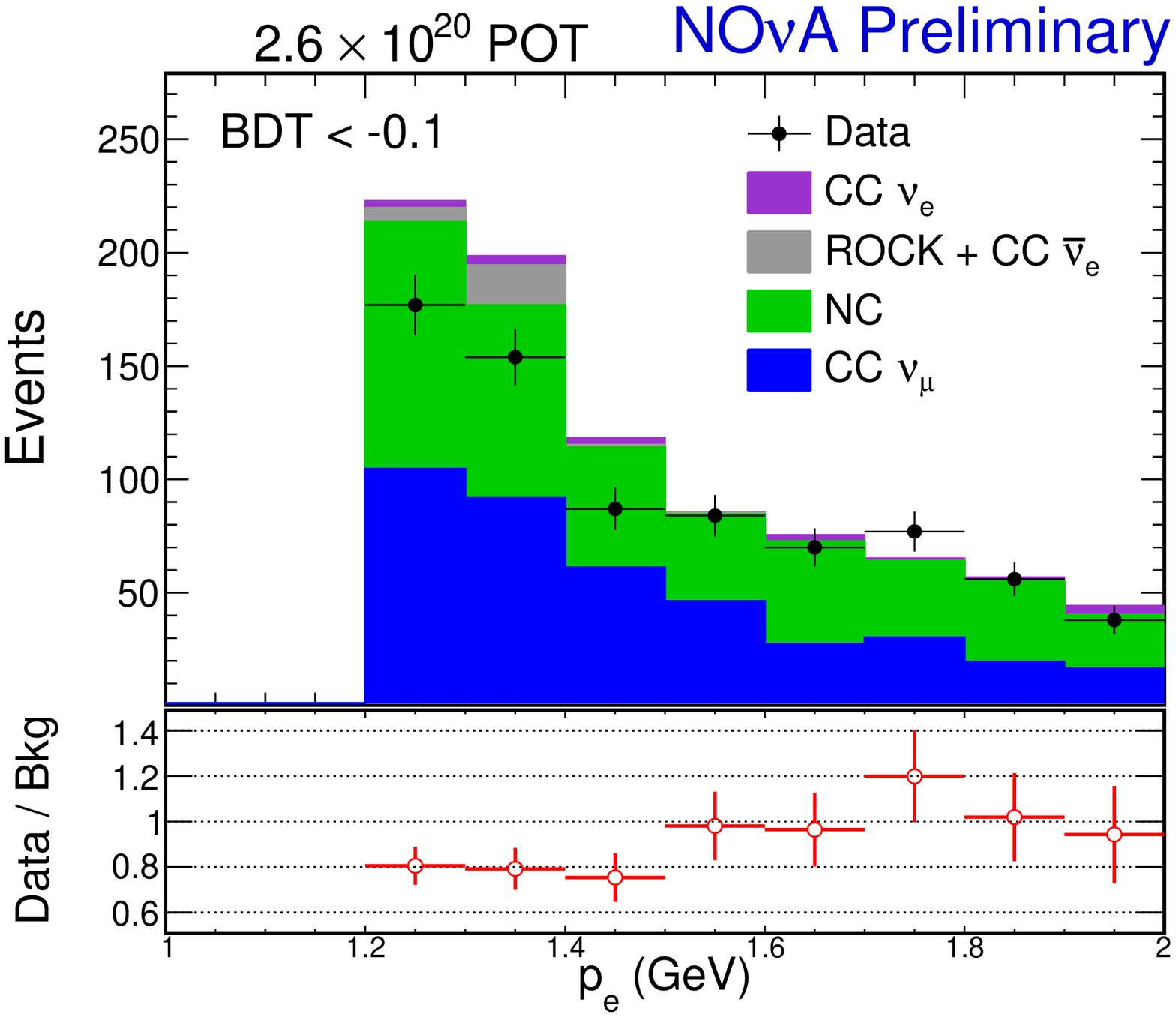}
\includegraphics[keepaspectratio,width=.4\textwidth]{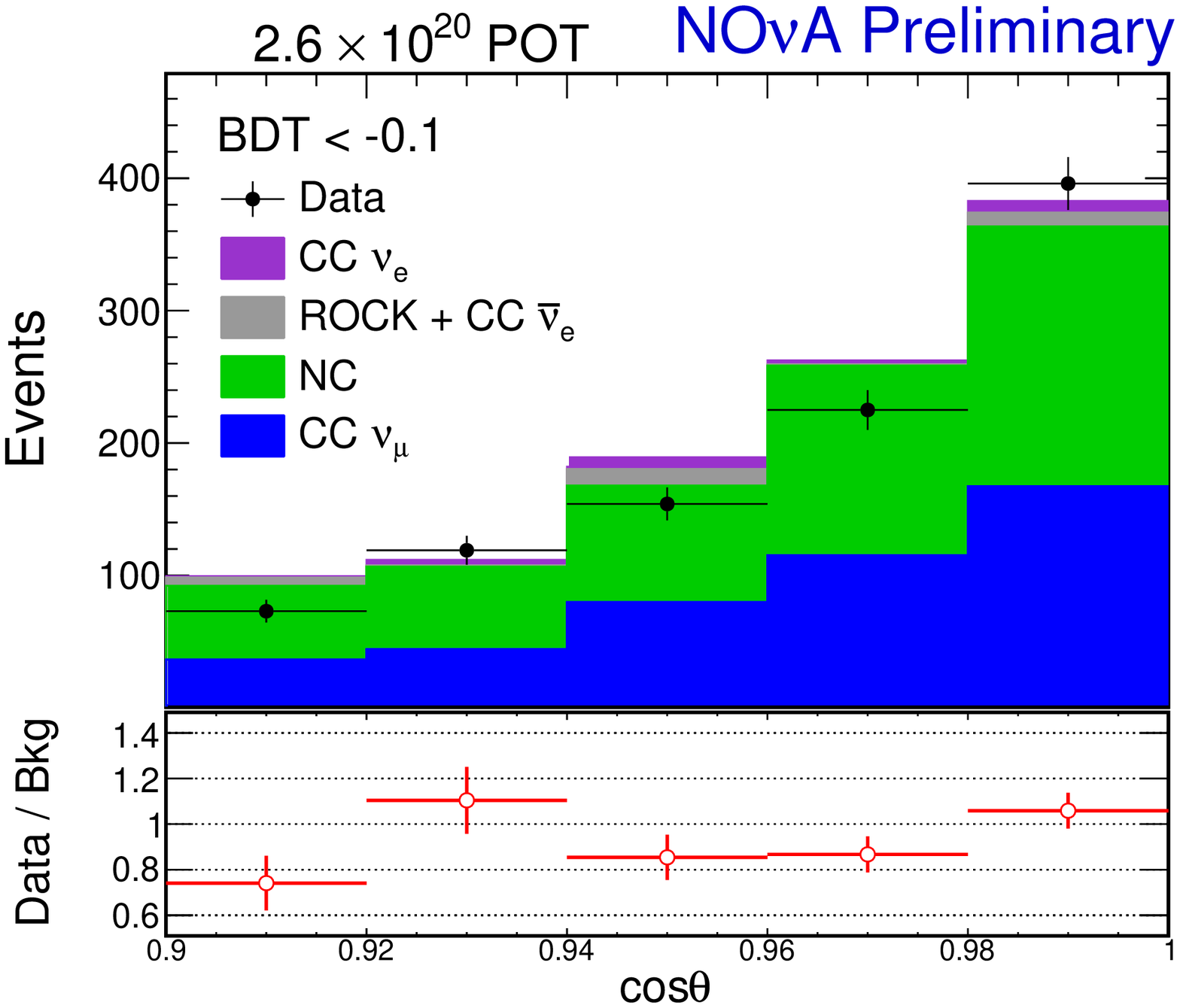}
\includegraphics[keepaspectratio,width=.4\textwidth]{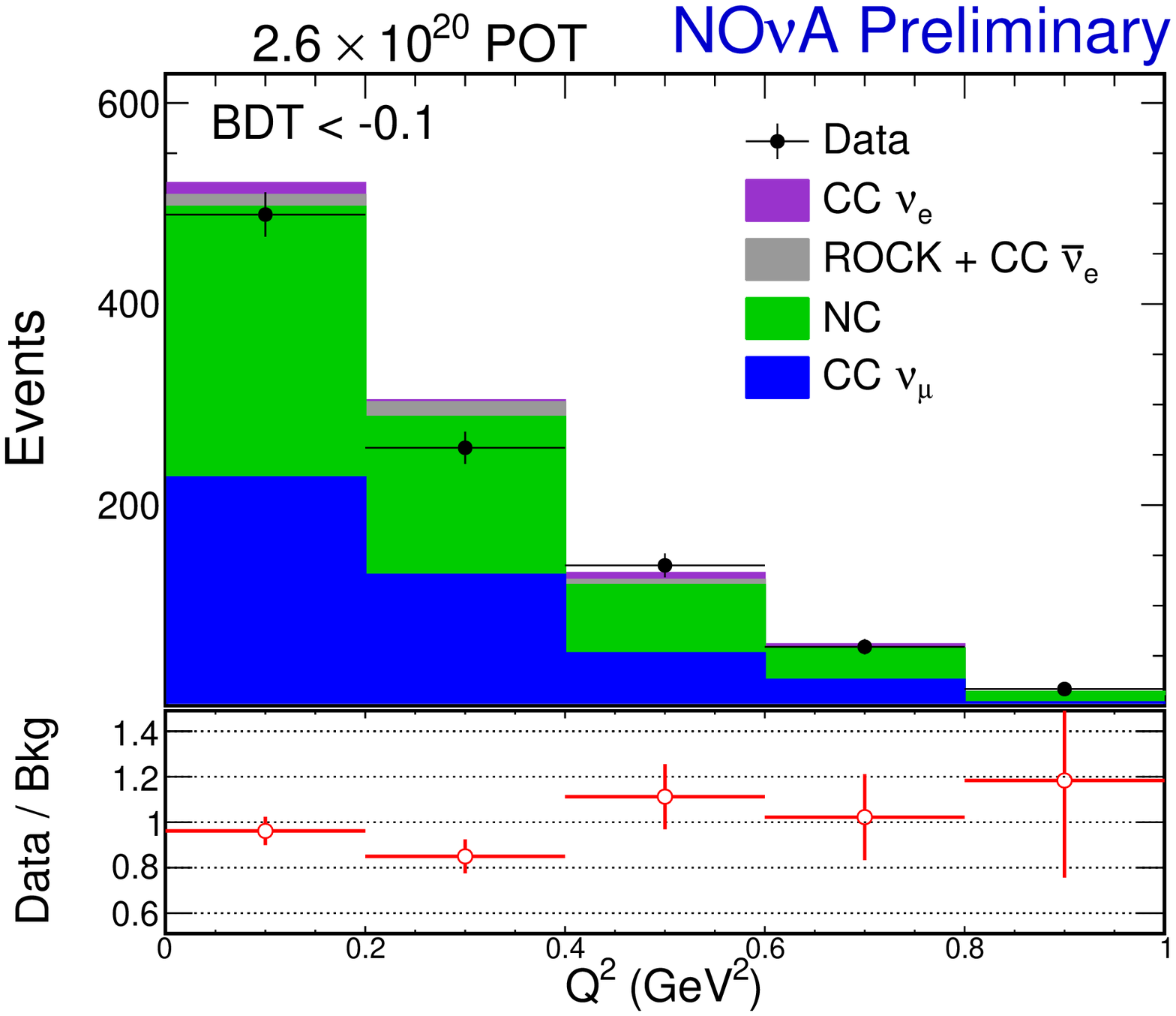}
\caption{Event distributions of reconstructed energy, primary shower momentum,
primary shower scattering angle respect to beam direction, and invariant 4-momentum
transfer squared from sideband samples in low BDT output region 
for data, signal, and backgrounds.}
\label{sideband-bdt}
\end{figure}

\begin{figure}[tbh]
\centering
\includegraphics[keepaspectratio,width=.4\textwidth]{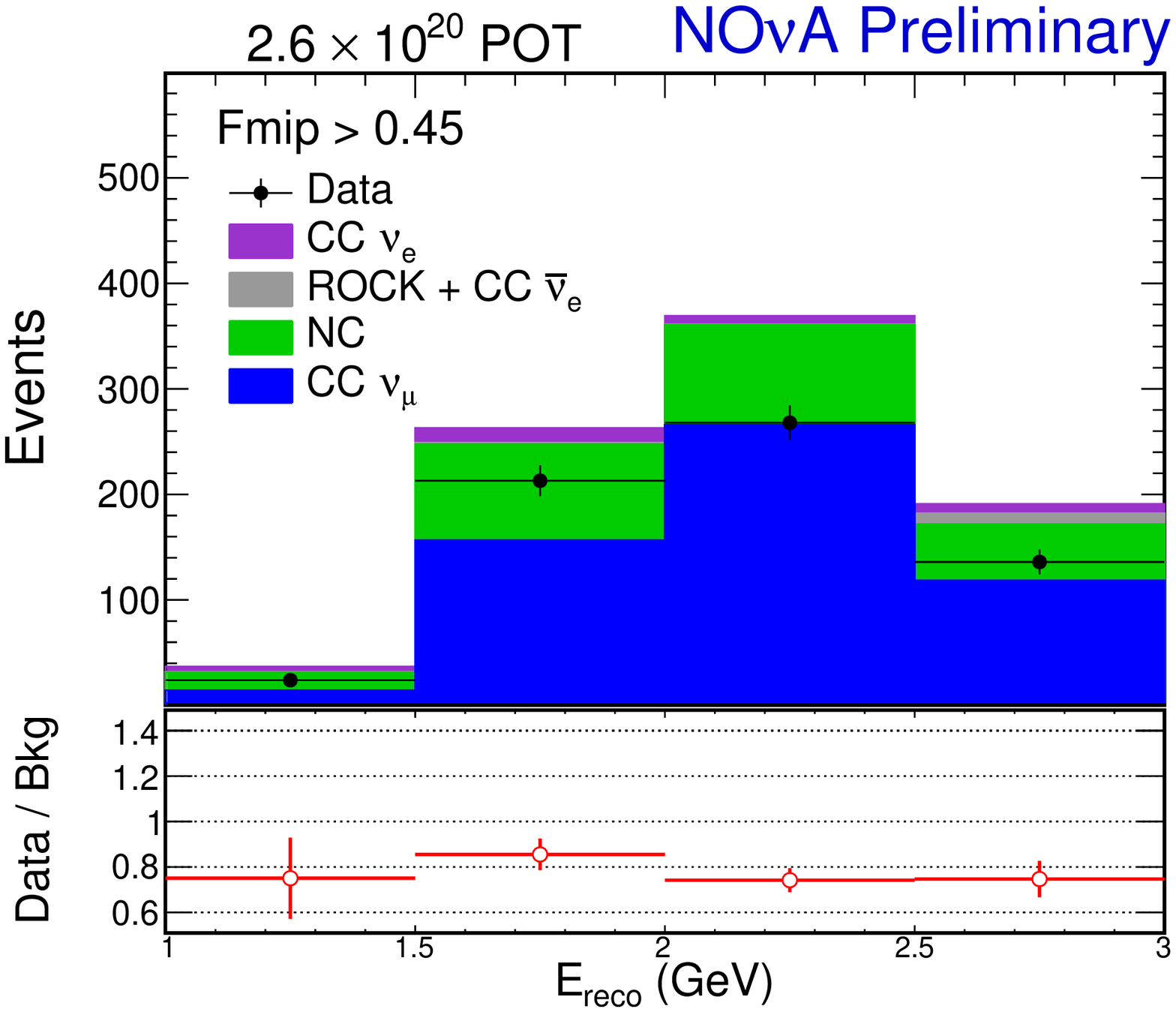}
\includegraphics[keepaspectratio,width=.4\textwidth]{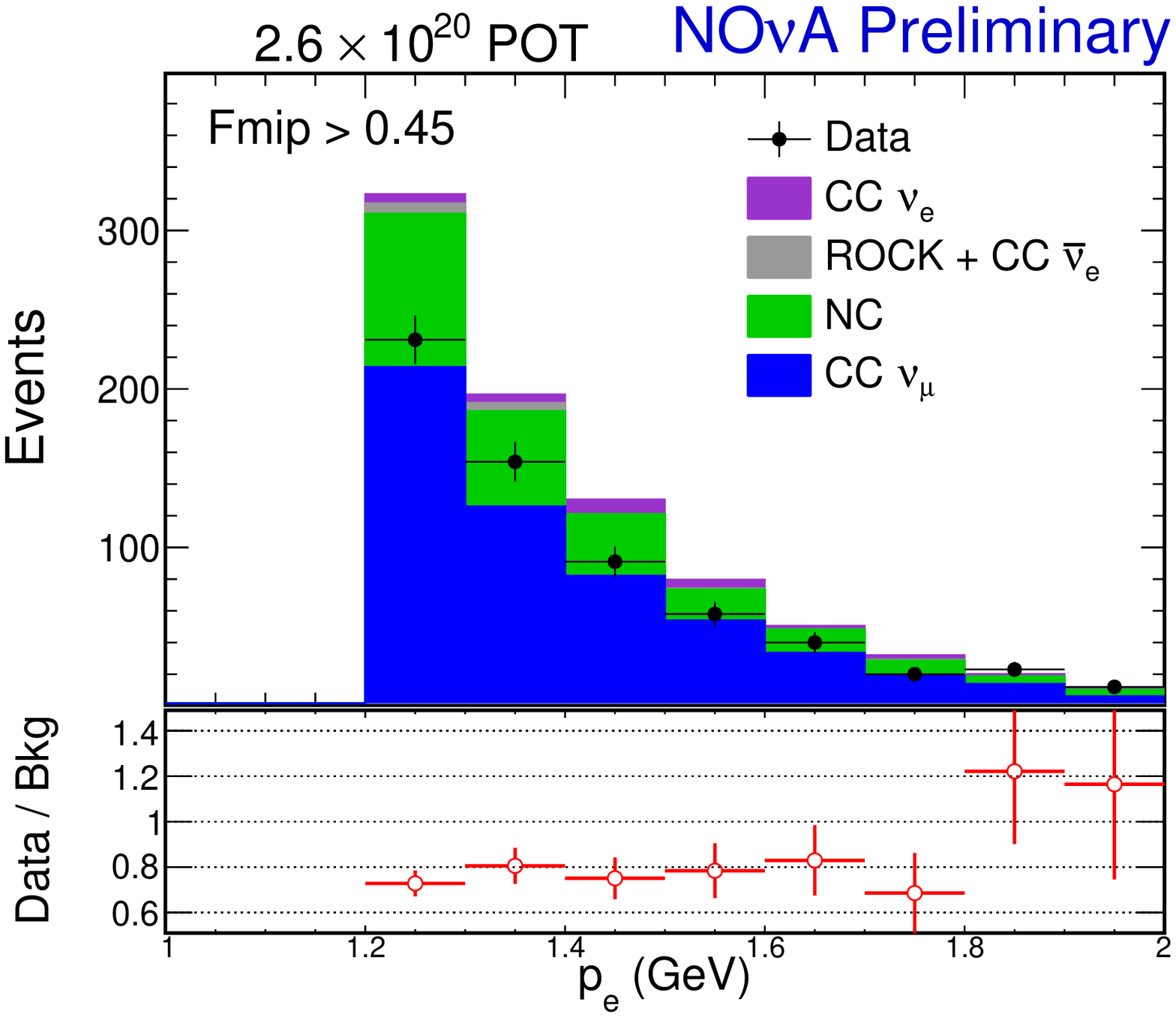}
\includegraphics[keepaspectratio,width=.4\textwidth]{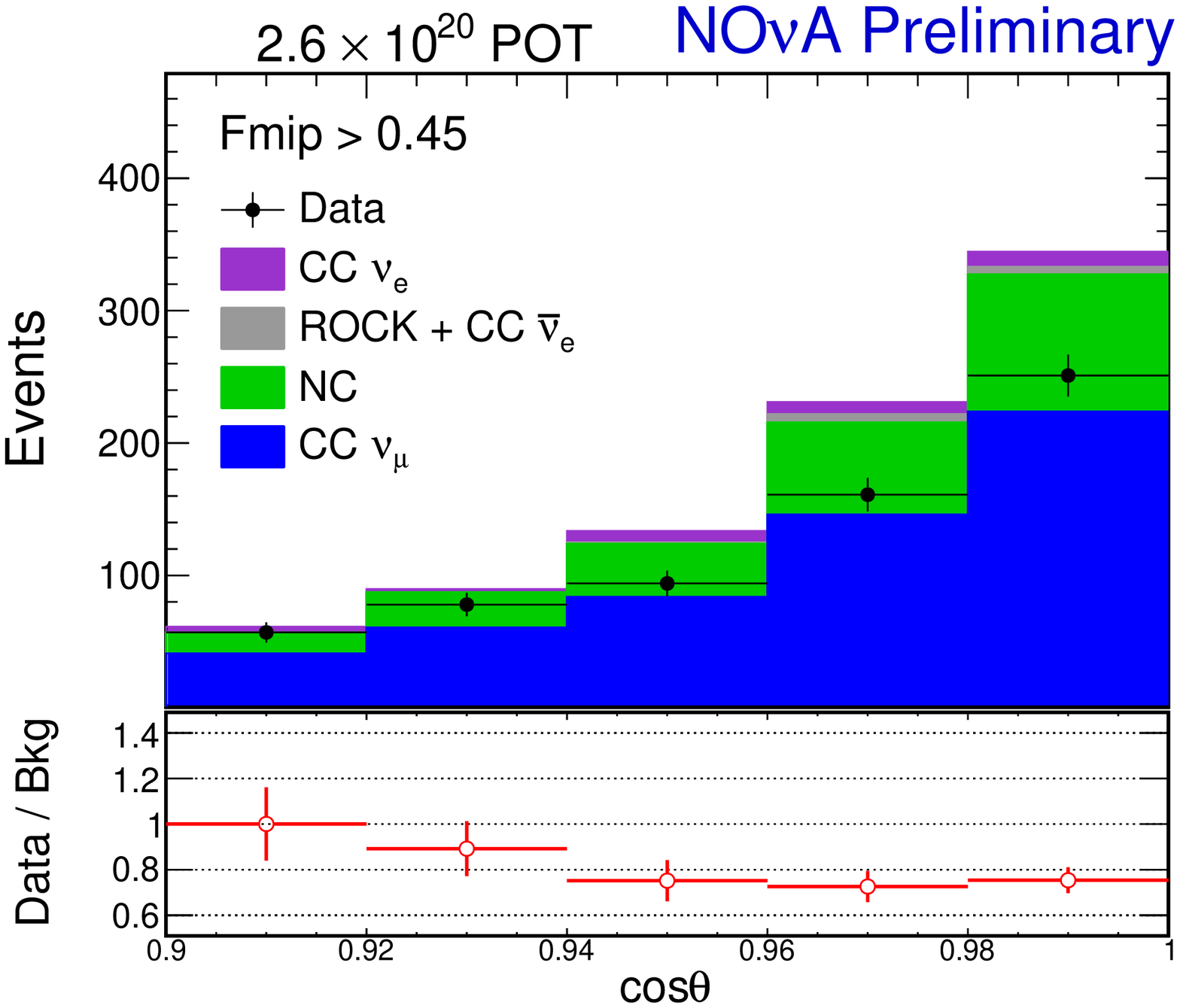}
\includegraphics[keepaspectratio,width=.4\textwidth]{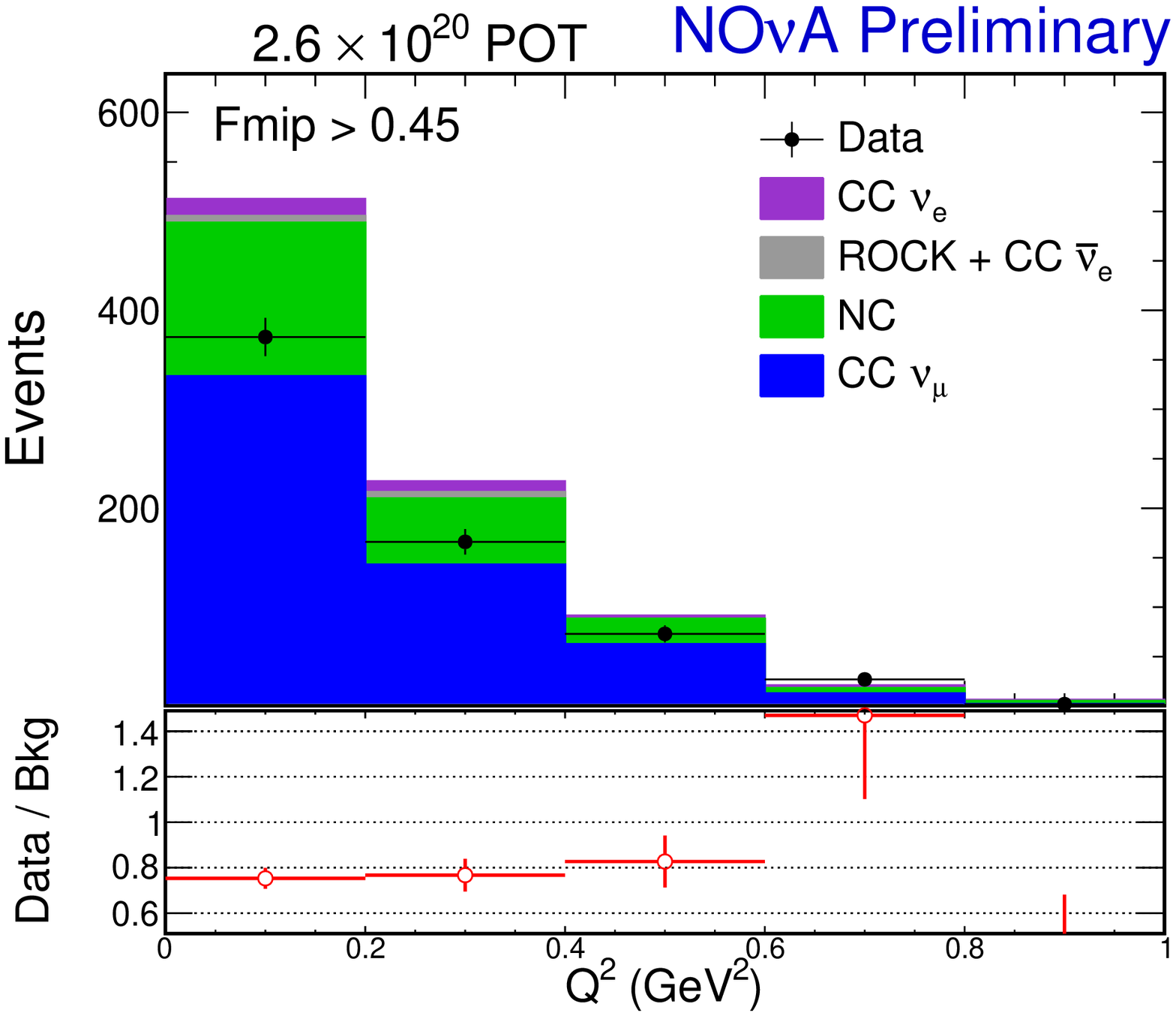}
\caption{Event distributions of reconstructed energy, primary shower momentum,
primary shower scattering angle respect to beam direction, and invariant 4-momentum
transfer squared from sideband samples with reversing the fraction of MIP hits
for data, signal, and backgrounds.}
\label{sideband-Fmip}
\end{figure}

The $\nu_{\mu}$ CC (including $\bar{\nu}_{\mu}$ CC),
NC, and $\bar{\nu}_{e}$ CC are the dominant backgrounds.
We explicitly treat the neutrino interactions outside the detector as the rock background,
which is dominated by $\nu_{\mu}$ CC events.
These backgrounds are estimated from {\sc genie} MC, 
and we derive the normalization factors from background dominant sideband samples to eliminate
the uncertainties from flux and {\sc genie} cross section modelings.
Two sideband samples are selected by either reversing the fraction of MIP cut on the primary shower,
or in the low BDT output region.
To select the events in the similar kinematic regions as the backgrounds in the
signal region, we further require the primary shower momentum to be greater than 1.2 GeV
and cos$\theta$ of primary shower to be greater than 0.9.
The sideband sample in the low BDT output region is used as the main sideband sample,
as it has similar ratio between NC and $\nu_{\mu}$ CC backgrounds as in the signal region.
The distributions for the reconstructed energy, primary shower momentum,
primary shower scattering angle respect to beam direction, and invariant 4-momentum
transfer squared for the two sideband samples are shown in 
Figs.~\ref{sideband-bdt}~and~\ref{sideband-Fmip}.
There is a clear normalization deficit between data and sum of the backgrounds.
Conservatively, we use 0.95~$\pm$~0.2 as the background normalization factor
to cover the bin by bin variation derived from the main sideband sample
with low BDT output, while also covering the difference between the two sideband samples.

\section{Systematics}
The dominant sources of systematic uncertainties are from
flux prediction, background normalization, hadronic energy,
and event selection efficiency.
There are two dominant uncertainties for flux prediction:
one is from the beam transport due to the horn current, beam spot size,
and magnetic field; the other is from the hadron production.
We use the external data results, including the pion and kaon production
cross section measurements from NA49 experiment~\cite{na49-pion,na49-kaon}, the kaon and pion ratio
using thin target from MIPP experiment~\cite{mipp-ratio}, and the recent pion yield measurements
based on thick target from MIPP experiment\cite{mipp-pion}.
Based on these external constrains, we reduce our $\nu_{e}$ flux by 5$-$8\%
from 1~GeV to 3~GeV energy region, and corresponding uncertainty is $\approx$10\%.
There is $\approx$21\% shape discrepancy for hadron energy between data and MC,
measured from $\nu_{\mu}$ disappearance analysis.
To take into account that discrepancy, 
we shift the hadron energy for $\nu_{e}$ CC signal by 21\% event by event
and the observed 6$-$10\% change on neutrino energy is used as the systematic uncertainty.
We cross-checked with selected $\nu_{e}$ CC candidates in sideband samples,
and confirmed the used 21\% shift is large enough to cover the existed discrepancy
for hadron energy between data and MC. 
Our event selection requirements including the BDT discriminant are dominantly based on the 
primary shower to be as model independent as possible.
The event selection efficiency is measured using {\sc genie} $\nu_{e}$ CC events.
The shower selection is studied using the bremsstrahlung showers in data and MC,
5\% uncertainty is obtained by comparing the selection efficiency as a function
of shower energy.
There is 5\% {\sc genie} cross section uncertainty, 
dominantly from CC Quasi-Elastic normalization.
The detector modeling effects have also been studied by using
different {\sc geant}4 physics lists, QGSP, QGSC, FTFP,
which is found to be negligible.

The statistical and systematic uncertainties for each energy bin
are summarized in Table~\ref{uncertainty}.
There is $\approx$20\% systematic uncertainty across energy region,
and we are already dominated by systematic uncertainty.

\begin{table}[tbh]
\centering
\caption{Statistical and systematic uncertainties.}
\captionsetup{justification=centering}
\label{uncertainty}
\begin{tabular}{cccc}
\hline
Energy (GeV) & Statistical Uncertainty (\%)& Systematic Uncertainty (\%)\\
1 $-$ 1.5 &22.6 & 22.4\\ 
1.5 $-$ 2 &11.8 & 20.4\\ 
2 $-$ 2.5 & 10.3& 21.5\\ 
2.5 $-$ 3 & 6.8& 18.7\\ 
\hline
\end{tabular}
\end{table}

\section{Results}
\begin{figure}[tbh]
\centering
\includegraphics[keepaspectratio,width=.4\textwidth]{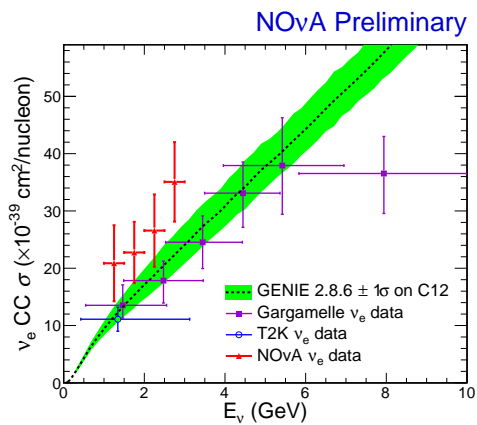}
\caption{The $\nu_{e}$ CC inclusive cross section in bins of neutrino energy.
The NO$\nu$A data points are shown in red triangle. The vertical errors represent
the total uncertainty, and the horizontal bar shows the energy coverage.
The {\sc genie} prediction for $\nu_{e}$ on C12 only is shown in dashed line,
with green band showing the 1$\sigma$ uncertainty.
The results from Gargamelle and T2K are also shown.
}
\label{xsection}
\end{figure}

\begin{figure}[tbh]
\centering
\includegraphics[keepaspectratio,width=.4\textwidth]{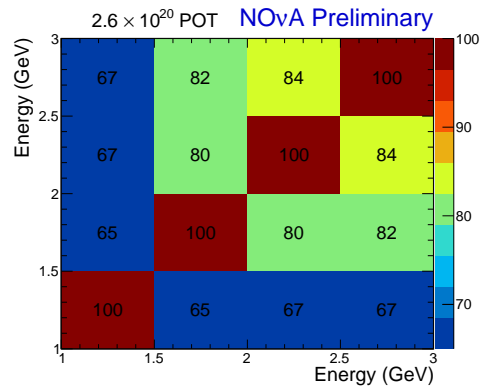}
\caption{Bin to bin correlation matrix.
}
\label{correlation}
\end{figure}

The measured $\nu_{e}$ CC inclusive cross section per nucleon
in 4 energy bins from 1~GeV to 3~GeV are shown in Fig.~\ref{xsection}.
The uncertainties are highly correlated between the 4 energy bins (see Fig.~\ref{correlation}).
The {\sc genie} prediction for $\nu_{e}$ on carbon only is also shown.

\end{document}